\documentstyle[aps,prl,epsfig,multicol]{revtex}
\begin{document}
\def\vp{\varphi}
\def\e{\epsilon}
\def\b{\beta}
\def\tG{\tilde{G}}
\def\heta{\hat{\eta}}
\def\hx{\hat{x}}
\def\hy{\hat{y}}
\def\hz{\hat{z}}
\def\vPhi{\vec{\Phi}}
\def\S{{\bf S}}
\def\r{{\bf r}}
\def\a{{\bf a}}
\def\k{{\bf k}}
\def\q{{\bf q}}
\def\l{\ell}
\def\j{{\bf j}}
\def\n{{\bf n}}
\def\vrho{{\vec{\rho}\,}}
\def\K{{\bf K}}
\def\J{{\bf J}}
\def\Q{{\bf Q}}
\def\E{{\bf E}}
\def\G{{\bf G}}
\def\B{{\bf B}}
\def\S{{\bf S}}
\def\R{{\bf R}}
\def\A{{\bf A}}
\def\X{{\bf X}}
\def\F{{\bf F}}
\def\M{{\bf M}}
\def\m{{\bf m}}
\def\T{{\cal T}}
\def\O{{\cal O}}
\def\H{{\bf H}}
\def\L{\Lambda}
\def\la{\langle}
\def\ra{\rangle}
\def\be{\begin{equation}}
\def\ee{\end{equation}}
\def\bea{\begin{eqnarray}}
\def\eea{\end{eqnarray}}
\draft
\title{Enhanced Pinning of Vortices in Thin Film
Superconductors by Magnetic Dot Arrays}
\author{R.~\v{S}\'{a}\v{s}ik$^*$ and T.~Hwa$^\dag$}
\address{
$^*$Institute for Pure and Applied Physics, University of California
at San Diego, 9500 Gilman Drive, La Jolla, CA 92093-0360\\
$^\dag$Department of Physics, University of California
at San Diego, 9500 Gilman Drive, La Jolla, CA 92093-0319
}
\date{\today}
\maketitle
\begin{abstract}
We study the pinning of vortices in thin film superconductors 
by magnetic dots in the London approximation. 
A single dot is in general able to pin {\it multiple} 
field-induced vortices, 
up to a saturation number $n_s$, which can be much larger than one.
However, the magnetic field of the dot also creates {\it intrinsic} vortices
and anti-vortices, which must be accounted for. 
In a ferromagnetic dot
array, the intrinsic anti-vortices are pinned only interstitially. Much stronger
pinning effect is expected of an antiferromagnetic dot array.
Possible realizations of various magnetic configurations are discussed.
\end{abstract}
\begin{multicols}{2}
\narrowtext

Artificially patterned sub-micron magnetic structures hold a great promise 
not only for magnetic device and storage technology, but also as 
a tool of fundamental research when used as a means of 
controlling other physical systems, such as the two-dimensional electron
gas~\cite{2deg} and vortices in superconductors~\cite{martin,bruyn,axel}. 
It is an experimental fact that when a type-II 
superconducting film is deposited over
a regular array of magnetic dots, resistivity
of the film exhibits a series of sharp minima as a function of the applied
magnetic field, with the positions of the minima being  integer multiples
of the geometrical ``matching field'' of the magnetic lattice~\cite{martin}.
In contrast, no periodic pinning was found in
a similar system with a regular array of {\em non-magnetic}
defects~\cite{martin}. This suggests the importance of the magnetization of the
dots in producing low resistivity, 
although the exact flux pinning mechanism has not been fully understood. 
A recent study by Lyuksyutov and Pokrovsky~\cite{pokrovsky} 
explored various statistical mechanics issues that might arise from the 
complexity of the magnet-superconductor interaction.
In this work, we focus on the pinning mechanism itself, the understanding
of which will lead us to desirable magnetic dot
structures that enhance vortex pinning for a range of applied magnetic fields.

We study the low temperature properties of a superconducting thin film
such as Nb deposited on top of a regular array of magnetic dots, separated
by a thin insulating layer to suppress the proximity effect (Fig.~1).
The superconductor has a magnetic penetration length
$\lambda$ much larger than the coherence length $\xi$,
and the film is
taken to be a homogeneous thin plate of thickness $d\ll\lambda$.
For the magnetic dots, we assume
the  magnetization on each dot to be {\em quenched in}, and pointing normal
to the layer.
In what follows, we will investigate how the interaction between vortices
in the superconductor is affected by the magnetic dots, and
explore which quenched configurations of magnetization
are favorable for vortex pinning at low temperatures. 
At the end, we will discuss how the
desired configuration(s) of magnetization might be achieved in practice.
\begin{figure}
\psfig{file=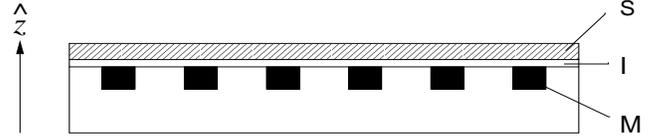,width=84mm,height=18mm}
\vspace{12pt}
\caption{
Schematic diagram of the superconducting layer (S) deposited over an array
of magnetic dots (M), separated with a thin insulating layer (I).}
\end{figure}

We first describe the case of a {\em single} dot with magnetization
$\M(\r)$ to be specified shortly.
Modeling the thin-film superconductor as an ideal sheet current 
$\K_s(\vrho)$ at the $z=0$ plane (with $\vrho\equiv[x,y]$ denoting the
position vector), Maxwell's equation of magnetostatics becomes
\be
-\nabla^2\A(\r)=\frac{4\pi}{c}\K_s(\vrho)\delta(z)+4\pi\nabla\times\M(\r),
\label{maxwell}
\ee
in the gauge $\nabla\cdot\A=0$.  We will describe the superconductor in
the London approximation~\cite{tinkham}, which has
$\K_s(\vrho) = (c/4\pi\Lambda)\cdot[\vPhi(\vrho)-\A(\vrho)]$,
where $\Lambda\equiv\lambda^2/d$ is the relevant magnetic length scale 
for the superconducting film, and $\vPhi(\vrho)$ is the London vector which,
for a collection of vortices located at points 
$\vrho_j$ with respective quantization (charges)
$\e_j$, is
\be
\vPhi(\vrho) = \frac{\phi_0}{2\pi}\sum_{j} \e_j\frac{\hz\times(\vrho-\vrho_j)}{
(\vrho-\vrho_j)^2}.
\ee
The above form of $\K_s$ holds everywhere except at the vortex cores, 
which are normal regions of radius $\sim\xi$ around each singularity,
and for as long as $\hz\cdot\A(\vrho)=0$, which will be 
satisfied in our case. Using the superposition principle, we write 
$\A(\r) = \A_s(\r)+\A_m(\r)$,
where $\A_s$ is the magnetic vector potential due to the supercurrent,
and $\A_m$ is the vector potential due to the magnetic dot.
In what follows, we will model a magnetic dot by a {\em perfect dipole} of  
magnetic moment $\m\equiv m\hz$, placed at a distance $\ell$
below the superconducting plane and the origin. Thus,
$\A_m(\r)= m(\hz\times\r)/|\r+\l\hz|^3$.
The dipole model is exact, of course, only for homogeneous spherical magnetic dots.
Nevertheless, we will use this dipolar approximation throughout, in the hope 
that the results will give the correct order of magnitude for a variety of 
magnets whose magnetizations are  normal to the plane.

The free energy of the entire system
may be written as 
\bea
F&=&\frac{1}{2c}\int d^2\vrho\,\left[\frac{4\pi\L}{c}|\K_s(\vrho)|^2+\A_s(\vrho)\cdot\K_s(\vrho)\right]\nonumber  \\
&&-\left.\m\cdot\nabla\times\A_s\right|_{\r=-\l\hz}.
\eea
The term in $[\ldots]$ is the sum of the kinetic and magnetic field energies of the
supercurrent; the last term is the potential energy of the magnetic dipole
in the magnetic field of the supercurrent. 
Using Eq.~(\ref{maxwell}), we can integrate out the vector potential and
express the free energy completely in terms of the vortices in the superconductors.
We have $F=F_{vv}+F_{vm}$,
where
\be
F_{vv} = \frac{\phi_0^2}{16\pi^2\L}\sum_{i,j}
\e_i\e_jU(|\vrho_j-\vrho_i|), \label{F-vv}
\ee
is the vortex-vortex interaction energy (including the vortex self-energy),
with $U(\rho)$ being the Pearl potential~\cite{pearl} whose asymptotic behavior 
is
\be
U(\rho)\sim\left\{ \begin{array}{ll}
\ln(\L/\xi), &\rho\ll\xi,
\\
(1/2)\ln(\L/\rho),& 
\xi\ll\rho\ll\L, \\
\L/(2\rho),&
\rho\gg\L, 
\end{array} 
\right.
\ee
and 
\be
F_{vm} = \frac{\phi_0m}{\pi\L^2}\sum_j\e_j
V(|\vrho_j|),\label{F-vm}
\ee
is the vortex-magnet interaction, with
\bea
V(\rho) 
&\equiv& -\int_0^\infty d\kappa\,
\frac{\kappa e^{-\kappa \l/\L}}{2\kappa+1}J_0(\kappa \rho/\L)\nonumber \\
&\sim&
\left\{ \begin{array}{ll}
-\L/2\l, &\rho\ll\l\ll\L,\\
-\L/2\rho,&\l\ll \rho\ll \L,\\
-2(\L/\rho)^3,&\rho\gg \L.
\end{array} 
\right.
\eea

We will assume that $\rho\ll\Lambda$ 
and  $\l\sim\xi$, which correspond to the more interesting 
and experimentally realistic cases. 
A single magnetic dipole with $m>0$ is in general able to bind
more than one vortex. The attractive force on a vortex due to the magnet
is the strongest
at a distance $\sim\l$ from the magnet. Since the vortex-vortex repulsion
decays only logarithmically with distance, we conclude that all bound 
vortices are concentrated in an area of radius $\sim\l$ around the dot.
When their number is large (and it can be), they can be considered a 
single multiply quantized vortex, since $\l\sim\xi$ by assumption.

The maximum charge $n_s$ of the bound vortex can be found as follows:
Suppose the net vorticity of the sample (as enforced by the application of 
an external magnetic field) is $N>n_s$, and out of these $N$ vortices, there is
a vortex of charge $n$ at the origin above the dot, 
with the remaining $N-n$ 
vortices singly quantized and far removed from each other and from the origin.
The total free energy of such a system is simply
\be
F(n)= \frac{\phi_0^2}{16\pi^2\L}(n^2+N-n)\ln\frac{\L}{\xi}+\frac{\phi_0m}{\pi\L^2}
n V(0).
\ee
$n_s$ can then be identified as the value of $n$ which minimizes $F(n)$, 
with the result
\be
n_s=\mbox{int}\left[
\frac{1}{2}+\frac{4\pi m}{\phi_0\l\ln(\L/\xi)}\right],
\label{max}
\ee
where $\mbox{int}[x]$ denotes the nearest integer to $x$.
The numerical value of $n_s$, which we shall call hereafter the {\em saturation number},
may be much larger than 1. As an example,
let us consider a ``typical'' 
magnetic dot made of Co/Pt multilayer, with magnetization $\sim 500$ emu cm$^{-3}$~\cite{lc},
dot size $(0.25 \mu\/{\rm m})^2$ and height $h \sim 40$~nm.
Taking $\l$ to be $h/2$, we get $m/\l \sim 3 \phi_0$.
The factor $\ln(\Lambda/\xi)$ is given by the superconducting film
itself; for Nb thin films somewhat below $T_c$ 
with $\xi\sim 20$~nm, $\lambda/\xi\sim 15$ and thickness $d\sim\xi$
one has $\ln(\Lambda/\xi)\sim 5$.
In this case $n_s \approx 8$. This example illustrates 
that magnetic pinning may be many times more 
effective than pinning by material defects (e.g., holes) at the same density,
since the holes can directly bind only one vortex per site.
The ability of a single magnet to bind so many
vortices is perhaps the most desirable property of the magnetic dot system,
at least for the purpose of vortex pinning; this property has however not
been recognized previously~\cite{pokrovsky}. 

Extrapolating the above-described properties of a single dot to systems
with many dots, one might
naively conclude that an array of dots in the ferromagnetic configuration
would provide the strongest pinning. Indeed, if the applied magnetic field
is $n_s$ times the matching field $B_\phi$, then each dot will
bind $n_s$ field-induced vortices, a task not achievable by an array of
non-magnetic pins if $n_s$ is large. 
However, the ferromagnetic dot array suffers from a different
problem: It has difficulty pinning  at small applied fields, i.e., 
for $B \ll n_s B_\phi$, because of the appearance of
{\em intrinsic} vortices and anti-vortices created by the magnetic field of 
the magnets themselves. 

To understand the effect associated with  intrinsic vortices, let us 
again consider the single-dot system,
but now in the absence of any external magnetic field.
For weak magnets, no vortex is created in the superconducting film.
Vortices eventually appear in the film for sufficiently strong magnets.
This first happens when there is a {\it doubly} quantized vortex 
bound directly above the magnet, with two single anti-vortices straddling
it, each at a distance $\rho_0$  away~\cite{note}.
The anti-vortices must be present in a large film (of linear size $\gg \l$), 
since the net magnetic flux through the $x$-$y$ plane due to the magnet is 
zero. (This important fact was left out of the model in 
Ref.~\onlinecite{pokrovsky}.)
Annihilation of the anti-vortices with the nucleus 
is prevented by the short-range repulsion between the
anti-vortices and the magnet.
To find the onset of intrinsic vortices, we compute the free energy $F(\rho)$
obtained by applying the forms of the vortex-vortex and vortex-magnet interactions
[Eqs.\ (\ref{F-vv}) and (\ref{F-vm})] to the vortex-antivortex configuration 
described above. We take $\rho_0$ to be the value of $\rho$ which minimizes
$F(\rho)$, with the result
$\rho_0 = 16\pi m/(3\phi_0)$. Intrinsic vortices 
appear  when $F(\rho_0)<0$, since  the free energy
of the system without vortices is set at $F=0$ by definition.
In the experimentally relevant limit $\rho_0\gg\xi$, this occurs when the
magnetization reaches a critical value
$ m_c = \frac{3}{8\pi}\phi_0\l\ln(\Lambda/\xi).$
Using this expression for $m_c$, we can rewrite Eq.~(\ref{max}) as
$n_s = {\rm int}[\frac{1}{2}+ \frac{3m}{2m_c}]$. Note that $n_s=2$
when $m=m_c$. Thus, intrinsic vortices and anti-vortices will appear 
once the magnet becomes a better pinning site than a simple
void, i.e., for $n_s > 1$.

Next, let us consider the behavior of a {\em ferromagnetic} (FM) square array
of dots with $m > m_c$ and lattice constant $a\sim\rho_0$,
again in the absence of any external
magnetic field.
In the Co/Pt example mentioned above, $a \sim \rho_0 \sim 1 \mu{\rm m}$.
In this case, the intrinsic anti-vortices
are very loosely associated with the magnetic dots and 
form a classical plasma, which interacts with the
FM array  only  interstitially.
For $n_s\gg 1$, the interstitial pinning is of high order 
and therefore very  weak. Consequently,
the anti-vortices can be set into motion by a small
applied current or thermal excitation, leading to dissipation 
even in the absence of any applied field!
This is certainly not a desirable feature for the purpose of 
vortex pinning.

The pinning ability of the FM array increases for 
increasing external fields, whose effect is mainly to annihilate
a fraction of the intrinsic anti-vortices at interstitial sites. 
The annihilation is complete
when the applied field reaches the order of $n_s B_\phi$ as already
described above. Thus maximum critical current is obtained at this
field. Through this consideration, we see that the main effect produced
by the ferromagnetic array is to {\em shift} the zero of the magnetic
field to some large value, i.e., $n_s B_\phi$. This can be
utilized in special applications which requires a high critical
current for a {\em given} field, but is not good in general, where high critical
current is demanded for a {\em range} of external fields.

The pinning properties of the sample in low applied fields can be 
significantly improved if the magnets form a quenched square
{\em antiferromagnetic} (AFM) array, 
where the intrinsic anti-vortices produced by those magnets in the $+\hat{z}$
(or ``up'') direction tend to be attracted to the magnets in the $-\hat{z}$ 
(or ``down'') direction. 
For strong magnets whose binding distance $\rho_0$ (between the 
magnet and its satellite anti-vortices) is of the order $a$, 
the anti-vortices will be pinned strongly to magnets pointed in the
opposite direction, leading to a much larger critical current needed
to dislodge the intrinsic vortices and anti-vortices at low fields.

In the absence of any external field, the 
free energy per unit cell of an
AFM square array is obtained from Eqs.~(\ref{F-vv}) and (\ref{F-vm}),
assuming that 
$n$ intrinsic vortices are localized on each upward magnets 
and $n$ intrinsic anti-vortices are localized on each downward magnet. 
To  sum the resulting alternating series of potential energies,
a good approximation is to take the two largest terms of the series, 
which  has a minimum when $n = n_s$. 

In the presence of an external magnetic field, the field-induced vortices
are subject to a periodic potential landscape created by the magnets and the 
bound intrinsic vortices and anti-vortices. 
To characterize this potential, we place a ``test'' vortex of charge $+1$
at some position $\vrho$, and compute the free energy $W(\vrho)$ experienced
by this test vortex, due to interaction with all the intrinsic vortices
and anti-vortices, $n_{i,j} = (-1)^{i+j} n_s$, and 
with all the magnets, $m_{i,j} = (-1)^{i+j} m$, 
on the respective lattice sites $\vec{R}_{i,j}=a(i\hat{x}+j\hat{y})$.
The resulting expression 
\bea
W(\vrho)&=&
\frac{\phi_0^2}{8\pi^2\Lambda} \sum_{i,j} n_{i,j} U(|\vrho-\vec{R}_{i,j}|)\nonumber\\
& &\qquad+ \frac{\phi_0}{\pi\Lambda^2} \sum_{i,j} m_{i,j} V(|\vrho-\vec{R}_{i,j}|), 
\label{W}
\eea
is plotted in Fig.~2 for the example in the text with $n_s=8$ and $a=\rho_0/2$.

\begin{figure}
\psfig{file=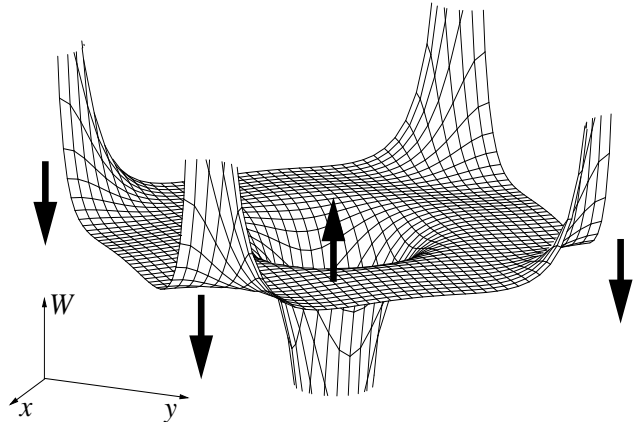,width=84mm,height=56mm}
\caption{
Free energy $W(\vrho)$ of a test vortex in an AFM dot array. 
The range of $W(\vrho)$ is $\pm \phi_0^2n_s/(16\pi^2\Lambda)$;
$\vrho$ spans the 
primitive cell 
of the AFM lattice. 
The arrows indicate magnetic moments of the dots on the lattice sites.}
\end{figure}

The form of $W(\vrho)$ clearly indicates that the test
vortex will be attracted towards the nearest ``up'' magnet.
Note that even though a magnet is already saturated with 
intrinsic vortices, it can still bind more field-induced vortices
because there is no core energy cost associated with the latter.
The maximum number of field-induced vortices $n_c$ 
that an upward magnet can bind is estimated as follows:  
Every field-induced vortex (per primitive cell)
that is attracted to the upward magnet changes 
the potential landscape for the test vortex.
These bound vortices surround the nucleus, effectively
increasing its charge.
A test vortex will no longer tend to the magnet when the repulsion
from the effective charge of the nucleus 
becomes equal to the maximum attractive force due to the magnet. 
The latter force is strongest at a finite distance $D$;
in the point dipole approximation, Eq.\ (\ref{F-vm}), 
$D\sim\l$. More realistically,
$D$ will be of order of the radius of the magnetic dot.
The balance of forces gives 
\be
n_c \approx n_s
\left(\frac{2\l}{D}\ln\frac{\Lambda}{\xi}-1\right).
\ee
In our example system, where $D\sim 0.1$~$\mu$\/m, $n_c\sim n_s$.

The analysis described in this study suggest very different behaviors for 
a superconducting thin film on a (quenched) 
FM or AFM dot array. For the FM array, we expect 
the system to have low critical current $J_c$ at low fields, with
successively increasing
$J_c$ when the applied field is an integer multiple of the matching field,
and with the maximal $J_c$ obtained  at $B=n_s B_\phi$. For the AFM array,
we expect the pinning to be the strongest at low fields, with successively
lower $J_c$ as the applied field reaches higher and higher orders of the
matching field, and with the main effect diminishing beyond a field value of 
$n_c B_\phi/2$. Thus, for different applications, one might want to use one
or the other type of arrays, or some combination.

It remains to address how the quenched magnetic dot configurations can be 
achieved and maintained in an applied field.
The FM array is most easily prepared by aligning the magnetic moments
in a strong magnetic field in the $z$ direction prior to measurement. It is
less straightforward to ensure the AFM arrangement.
Fortunately, we are aided in this
case by the fact that the AFM state is 
the {\em ground state} of a square lattice of
magnetic dipoles as long as the out-of-plane direction (i.e., the $\hat{z}$ axis)
is the ``easy'' magnetization axis\cite{mac}. The latter can be arranged
by the construction of the individual magnetic dot, e.g., by using the multi-layer
Co/Pt structure\cite{vanbael}. The appearance of intrinsic vortices further 
stabilizes the AFM structure.
Thus, the AFM array will in fact form 
{\em spontaneously} upon cooling from high temperature in zero applied field, 
based on purely energetic considerations.
There are, of course,
kinetic constraints, such as coercive effects and the mobility of the 
domain wall separating the two degenerate states 
of the antiferromagnet, that may prevent a perfect antiferromagnet from being
formed at a reasonable time.  These kinetic constraints are, on the other hand,
necessary to prevent the magnets to undergo a spin-flop type transition~\cite{flop}
to the FM phase upon increasing the external magnetic field.
One can also envision building an array of microscopic superconducting 
current rings below the film, and electronically
control the sense of current in each ring. This way, arbitrary quenched
magnet configurations can be specified. 
\begin{figure}
\psfig{file=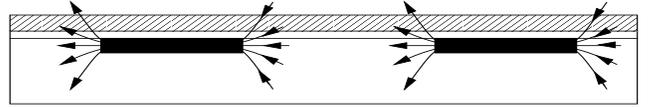,width=84mm,height=14mm}
\vspace{12pt}
\caption{
Schematic diagram of the superconducting layer 
with thin magnetic rods magnetized horizontally.}
\end{figure}

A practical way to achieve effects similar to that of the AFM array
is to bury under the superconductor an array of
thin magnetic {\em bars} 
(Fig.~3). The shape anisotropy of the bar
will force the magnetization to be along the length of the magnet. 
The field outside the magnet is the same as that of two oppositely charged
magnetic monopoles located at its endpoints,
creating an effect similar to that of
an antiferromagnetic pair of dipoles whose moments are normal to the 
sample~\cite{sh}.
A crude approximation to this geometry is the use of thin  magnetic disks
whose diameter is a significant fraction of the lattice constant. 
The anisotropic
shape of the disk induces a magnetization  parallel
to the superconducting plane, similar to that of the magnetic bar.
This was in fact the geometry used by Mart\'{\i}n 
{\it et al.}~\cite{martin}.

We gratefully acknowledge useful conversations with A. Hoffmann
and I. K. Schuller. This research is supported by the NSF through
grant no.\ DMR9801921, and by the UC-CLC program.

%
%

%
%
\end{multicols}

\begin{references}

\bibitem{2deg} D. Weiss, K. von Klitzing, K. Ploog, and G. Weimann,
Europhys. Lett. {\bf 8}, 179 (1989).

\bibitem{martin} J. I. Mart\'{\i}n, M. V\'{e}lez, J. Nogu\'{e}s, and 
I. K. Schuller, Phys. Rev. Lett. {\bf 79}, 1929 (1997).

\bibitem{bruyn} M. J. Van Bael, K. Temst, V. V. Moshchalkov, and Y. Bruynseraede, 
Phys. Rev. B {\bf 59}, 14674 (1998).

\bibitem{axel} J. I. Mart\'{\i}n, M. V\'{e}lez, A. Hoffmann, I. K. Schuller,
and J. L. Vincent, Phys. Rev. Lett. {\bf 83}, 1022 (1999).

\bibitem{pokrovsky} I. F. Lyukyutov and V. Pokrovsky, Phys. Rev. Lett.
{\bf 81}, 2344 (1998).

\bibitem{tinkham} M.~Tinkham, {\it Introduction to Superconductivity},
(McGraw-Hill, New York, 1996).

\bibitem{pearl} J. Pearl, Appl. Phys. Lett. {\bf 5}, 65 (1964).

\bibitem{lc} Z. G. Li and P. F. Carcia, J. Appl. Phys. {\bf 71}, 842 (1992).

\bibitem{note}
An isolated magnet with vertical magnetic moment 
cannot create a single vortex-antivortex
pair that would be stable with respect to annihilation; this statement
however does not apply when the magnet is part of an array.

\bibitem{mac} A. B. MacIsaac, K. De'Bell and J. P. Whitehead, Phys. Rev. Lett.
{\bf 77}, 739 (1996).

\bibitem{vanbael} M. J. Van Bael {\it et al.}, cond-mat/9911033 (preprint).

\bibitem{flop} J. M. Kosterlitz, D. R. Nelson and M. E. Fisher,
Phys. Rev. B {\bf 13}, 412 (1976).

\bibitem{sh} R. \v{S}\'{a}\v{s}ik and T. Hwa, in preparation.

\end{references}
\end{document}